\documentclass[twocolumn,pre,aps,10pt]{revtex4-1}

\usepackage{amsfonts,amssymb}
\usepackage[sumlimits,intlimits]{amsmath}
\usepackage{graphics}
\usepackage{graphicx}
\usepackage{epsfig}
\usepackage{mathrsfs}
\usepackage{wasysym}
\usepackage{textcomp}
\usepackage{verbatim}
\usepackage{hyperref}    
\usepackage{bm}          

\newcommand{\be}{\begin{equation}}
\newcommand{\ee}{\end{equation}}

\begin{document}

\title{Cooperative Charging in a Nanocrystal Assembly Gated By Ionic Liquid}

\author{Tianran Chen, Brian Skinner, B. I. Shklovskii}
\affiliation{Department of Physics, University of Minnesota, Minneapolis, Minnesota 55455}

\date{\today}

\begin{abstract}

In order to make a densely packed assembly of undoped semiconductor
nanocrystals conductive, it is usually gated by a room temperature ionic liquid.  The ionic liquid enters the pores of the super-crystal assembly under the influence of an applied voltage. We study the capacitance of such a device
as a function of the gate voltage. We show that,
counter-intuitively, the capacitance of the system is the sum of
delta-functions located at a sequence of critical gate voltages. At each critical
voltage every nanocrystal acquires one additional electron.

\end{abstract}
\maketitle

\section{Introduction}

Densely packed assemblies of monodisperse semiconductor
nanocrystals (NCs) are intensively studied as new materials with
tunable optical and electronic properties. They potentially can be
used for solar cells, light emitting diodes, transistors and
photodetectors~\cite{Murray, Efros}. When NCs are not doped and,
therefore, are insulating one can gate them
electrochemically~\cite{Yu}, for example, by a room temperature
ionic liquid (RTIL). Electrochemical gating allows one to bring
several electrons per NC~\cite{Heng}. It was shown that charge
transport through the assembly is characterized by variable range hopping, whose parameters change non-monotonically with the gate voltage $V$ \cite{Heng}. This calls
for an understanding of the charging process, namely a calculation of the
capacitance of a densely packed NC assembly as a
function of $V$, which is absent in the literature.

\begin{figure}[t!]
\centering
\includegraphics[width=0.4 \textwidth]{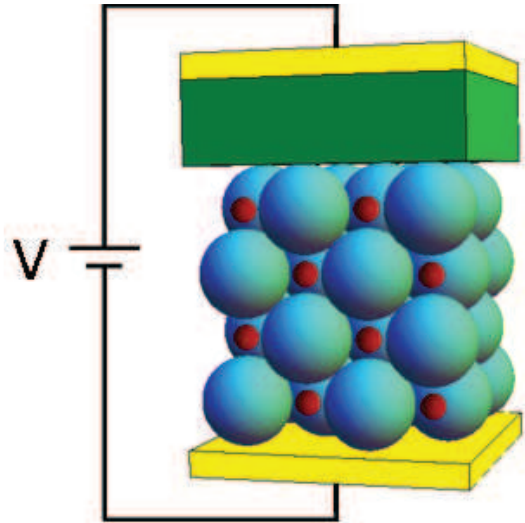}
\caption{A section of a three-dimensional face-centered cubic crystalline assembly (super-crystal) of semiconductor NCs (blue spheres, light gray) resting on a grounded metallic contact (yellow, very light gray). The SC is covered from above by a RTIL (green, gray) in which a second metallic contact (yellow, very light gray) is immersed. When a voltage $V$ is applied between the top and bottom metallic plates, cations from the RTIL (small red spheres, dark gray) fill the octahedral pores of the SC to compensate the negative charges of electrons arriving to NCs from the bottom electrode.} \label{fig:device}
\end{figure}

In this paper, we are concerned with the charging process of a
multi-layer, three-dimensional super-crystal (SC) of semiconductor NCs resting on
a grounded metallic contact (see
Fig.\ \ref{fig:device}). The SC is covered from above and gated by
a bulk RTIL in which a second metallic contact is immersed. The RTIL is a
good conductor, so that when the voltage $V$ is applied between the top and
bottom metallic contacts, anions from the RTIL are attracted to the top electrode and cations are attracted to NCs that are negatively charged by electrons. As a result, there are effectively
two series capacitors, one formed by the top
metal-RTIL contact and the other formed at the bottom RTIL-semiconductor SC contact. Both contacts are
blocking with respect to transport of ions from the RTIL and electrons into the
RTIL over a large window of $|V|$, so that no insulator is necessary.

\begin{figure}[t!]
\centering
\includegraphics[width=0.35 \textwidth]{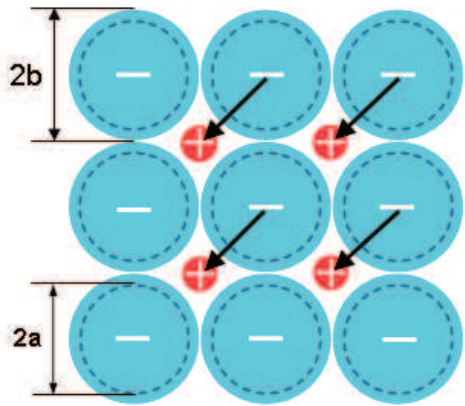}
\caption{Cross section of the face-centered cubic lattice of
NCs along the (100) plane. One can see the rock salt lattice of singly
charged NCs (large blue spheres, gray) and cations (small red
spheres, dark gray). The internal diameter $2a$ and external diameter $2b$ are shown. The arrows show a possible choice of
dipoles made from nearest neighbor cations and electrons.}
\label{fig:lattice}
\end{figure}

In a typical experiment the contact area of the top metallic
electrode is made so large (not shown in Fig.\ \ref{fig:device})
that the latter capacitance plays no role in the total capacitance between the two gating electrodes, since it is placed in series with the much smaller SC-RTIL capacitance. In other words, all
the external gate voltage $V$ drops at the SC-RTIL
capacitor, which is the object of our interest below. We assume
that all NCs are identical spherical particles of internal radius
$a$ and are covered by a thin layer ($\sim$ 0.5 nm) of an insulator
to prevent them from being sintered. We further assume
that the external diameter (including the insulating layer)
of NCs,  $2b \sim 5$ nm ($a$ and $b$ are shown in Fig.\ \ref{fig:lattice}), is
large enough so that both cations and anions of the RTIL, which have
diameter $\sim 1$ nm, can freely percolate from the bulk of the RTIL
into the pores between spheres of the SC and thereby establish unique, uniform electrochemical potentials. Thus, the RTIL
can be considered as an equipotential ionic conductor. It
effectively serves as one of the electrodes of the bottom
capacitor we would like to study. The other electrode is the
SC itself. 

We assume that the electrons of all NCs in the SC have the same
electrochemical potential as the bottom metallic electrode in
spite of the fact that the semiconductor NCs are covered
by an insulating layer. This implies that tunneling across the
barrier between
two neighboring NCs is fast enough that electrons always have
enough time to hop between NCs many times during the period of the AC
voltage signal used for capacitance measurement. In this sense, the bottom metallic electrode and the SC together constitute the equipotential electrode of the bottom capacitor.  Again, the RTIL is the second electrode. On the other hand, we also assume that the electron tunneling barrier between neighboring NCs is large
enough to allow only an integer number of electrons in each NC.
This requires the tunneling matrix element to be smaller than any
residual disorder distinguishing between neighboring NCs.

\section{Nontrivial character of charging of SC-RTIL capacitor}
\label{sec:}

Let us now consider charging of the SC-RTIL capacitor. We assume that semiconductor NCs are not doped, so that some gate
potential is necessary to align the Fermi level of electrons with
the conduction band of the semiconductor. In this paper we
calculate the gate voltage $V$ from this reference point. Our goal is to
calculate the three-dimensional concentration of electrons $n$ as
a function of $V$ as well as the volumetric differential capacitance
(capacitance per unit volume) of the SC: $C(V) = e dn/dV$. Of
course, the validity of this concept requires that, e.g., the capacitance of the SC
be proportional to the number of NC layers in the NC film.

\begin{figure}[t!]
\centering
\includegraphics[width=0.4 \textwidth]{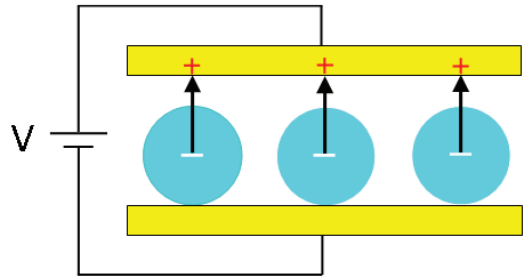}
\caption{An example of a more conventional capacitor. One layer of NCs (blue spheres, gray) is charged by parallel plane metallic gate (yellow, light gray).  Electrons within the NCs make dipoles with counter charges (arrows) in the top gate. All dipoles are collinear and reside in a single plane perpendicular to the two parallel metal plates. The repulsion between dipoles leads to monotonous growth of $n(V)$.}
\label{fig:standardcapacitor}
\end{figure}

It is natural to think that the charging of all NCs of the SC assembly happens sequentially with
growing $V$, i.e., NC by NC, because as in any
capacitor an electron just added to a particular NC repels other
electrons and prevents them from entering neighboring dots.
This indeed would be true if we were dealing with a single layer of
NC charged by parallel, planar metallic gate hanging above it and
separated by a thin insulator (Fig.\
\ref{fig:standardcapacitor}). In such a system each electron arriving to
a NC makes a dipole with its image in the top
gate. All dipoles reside in a single plane perpendicular to the two metal
plates. They, therefore, repel each
other and at small concentrations form a correlated liquid
of dipoles or dipole Wigner like crystal, which provides resistance to
charging of the NC layer. The capacitance of such a system was calculated
in Ref.~\cite{BrianS} and at finite temperature it is everywhere
positive and finite, so that $n(V)$ is a monotonically growing function.
In this sense, the system shown in Fig.\ \ref{fig:standardcapacitor} is qualitatively similar to a
conventional plane capacitor made from two metallic parallel electrodes,
the energy of which may be calculated as the result of pairwise repulsive
interactions between elementary, continuously distributed dipoles created by
positive and negative infinitesimal charges on two opposite capacitor plates.

\begin{figure}[t!]
\centering
\includegraphics[width=0.35 \textwidth]{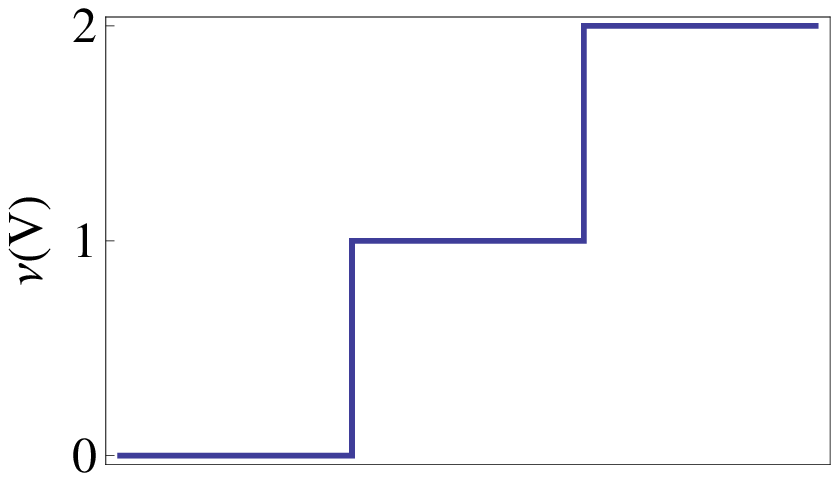}

\includegraphics[width=0.35 \textwidth]{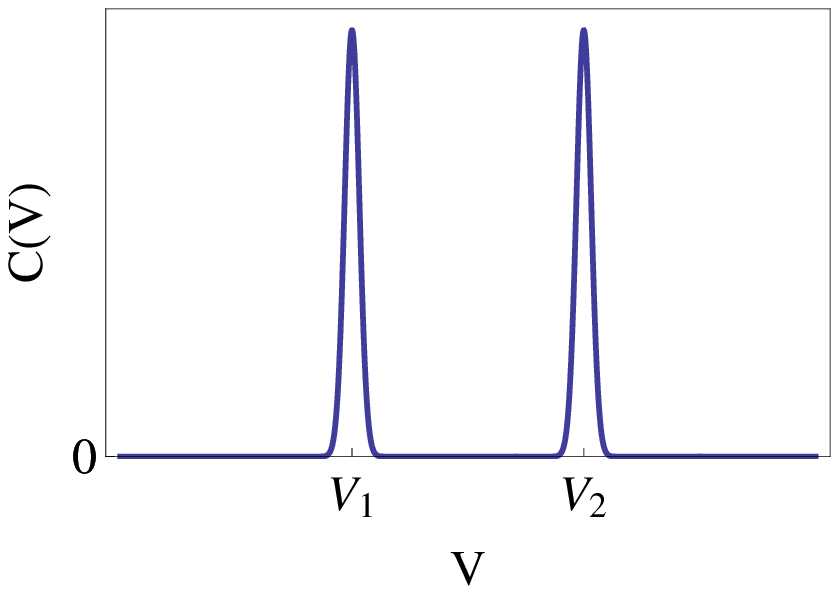}
\caption{Schematic plots of the filling factor of NCs
$\nu(V) =n(V)/N$ and the SC-RTIL capacitance as a function of 
voltage $V$. $\nu(V)$ shows alternating
long plateaus where it is an integer and vertical steps,
where $\nu(V)$ changes by unity. The
capacitance C is the sum of delta-functions located
at special voltages $V_k$, at which $k$-th electron enters each dot.}
\label{fig:charcap}
\end{figure}

In this paper, however, we predict that in a SC-RTIL capacitor
the electron density $n(V)$, or, equivalently, the filling factor of NCs $\nu(V) = n(V)/N$,
shows alternating long plateaus and vertical steps. At
each step $n(V)$ changes by the concentration of NCs $N$ or, in
other words, each NC of the SC acquires at the same $V$ exactly
one additional electron. This means that $C(V)$ is the sum of a
small number of delta-functions located at special voltages
$V_k$, at which the $k$-th electron enters $\it{each}$ dot. Schematic
plots of the filling factor $\nu$ and capacitance $C$ of the SC as a
function of gate voltage $V$ are shown in Fig.\ \ref{fig:charcap}.
Together, the whole SC behaves like a single quantum dot in spite of the interaction
between NCs. Indeed, in a single NC or a quantum dot the $k$-th electron is admitted
at the critical voltage $V_k = ke/C$, where $C$ is the gate-dot capacitance.
The voltage $V_k$ satisfies $e V_k = E_{v=k} - E_{v=k-1}$, where $E_{v=k}$ is the energy per NC in a SC with filling factor $\nu = k$.

Throughout this paper, we use the term ``cooperativity" to refer to
the fact that all NCs of the SC simultaneously acquire an extra electron at a particular voltage as a consequence of an attractive net interaction between electron-cation dipoles.
Thus, our goal in this paper is to prove and understand cooperativity.
We concentrate mostly on the case of the transition between $\nu = 0$ and
$\nu = 1$.

In Sec.\ \ref{sec:semi}, we study a simple model of semiconductor NCs with large Bohr radius $a_B$, i.e., weak Coulomb
interaction, and demonstrate the existence of cooperativity. In Sec.\ \ref{sec:metal}, metallic NCs are discussed and are proven to show cooperativity of electrons as well. Appendices A, B, and C deal with computational and mathematical details. 

\label{sec:intro}

\section{Semiconductor NC Assembly}

In this section, we focus on a densely packed semiconductor NC assembly.
Although in a real system the NCs usually have a different dielectric constant $\epsilon$ than that of the
RTIL, as a pedagogical example to illustrate cooperativity,
$\epsilon$ is for the moment taken to be uniform everywhere.
More realistic cases with non-uniform dielectric constant
will be discussed later in this paper.
We show below that the counter-intuitive behavior of the SC-RTIL
capacitor arises from the combination of two conditions: 1) the
electrodes of the SC-RTIL capacitor interpenetrate each other, and 2) charges of both signs are discrete and localized.

Let us assume that we deal with a 3D face-centered cubic
SC of densely packed NCs (Fig.\ \ref{fig:device}). Within an fcc lattice, there are two types of pores, octahedral and tetrahedral. (For visualization of
octahedral and tetrahedral pores see Ref.~\cite{crystal}.) There are two tetrahedral pores and one octahedral pore per NC. The octahedral pore can admit a sphere with radius $r = (\sqrt{2}-1)b\simeq 0.41b$, while a tetrahedral one can admit a sphere with radius $r = (\sqrt{6}/2-1)b\simeq 0.22b$. Let us first assume that NCs are relatively small and cations can only fit into the octahedral pores. In this case, cations can fill up all the octahedral pores of the SC in order to make the assembly overall neutral (Fig.\ \ref{fig:device}). Each cation
is assumed to be localized in the center of an octahedral pore. A single electron in a
NC is confined by the NC wall and is also affected by the Coulomb
potentials of cations and negative charges of other NCs. We
assume that the radius $a$ of the semiconductor NC is much smaller than the effective Bohr radius $a_B =
\hbar^2\epsilon/me^2$ of the semiconductor. Here $m$ is the
effective electron mass in the semiconductor. In this case, in the first
approximation the kinetic energy dominates the Coulomb energy and we may use the unperturbed
electron ground state wave function 
\be
\psi(r) = \sqrt{\frac{1}{2 \pi a}} \frac {\sin \left( \pi r/a
\right)}{r} 
\label{eq:wavefunction} 
\ee 
and the corresponding ground state
energy per NC 
\be 
E_q = \frac{\hbar^2 \pi^2}{2 m a^2}.
\label{eq:Eqm} 
\ee

For a calculation of the main term of the electrostatic energy of the
system, one can think that each electron is located in the center
of the NC. Octahedral pores form an fcc lattice that is shifted in space relative to the fcc lattice on which the NCs reside.  These two alternating fcc lattices, the one of singly
charged NCs and the one of cations, together form a rock-salt 
lattice of point charges. Because of complications of showing a
three-dimensional lattice, we illustrate a mutually penetrating
SC of NCs and the crystal of cations for the case of a 2D square rock-salt
lattice SC at $\nu = 1$ (Fig.\ \ref{fig:lattice}).

One may divide the rock salt crystal into neutral pairs of a cation
and an electron in neighboring NC and think that these pairs enter
sequentially as the voltage $V$ increases. However, the
following consideration shows that such pairs enter simultaneously
at some threshold voltage rather than sequentially. Let us
first calculate the voltage $V_s$ at which one such neutral pair enters an initially empty lattice.
Bringing one electron into a NC costs quantum mechanical energy
$E_{q}$, plus some energy $E_{ion}$ to bring a
cation into a pore of the empty system from the bulk of the RTIL. Generally, $E_{ion}$ is determined by the chemical potential of ions in the bulk RTIL, which is of the order of the nearest neighbor interaction between ions in the bulk RTIL. In addition, one should include
the Coulomb interaction between the electron and the cation: $ -e^2
/\epsilon \sqrt{2} b$. Thus, the total energy associated with bringing one electron-cation pair into the lattice is
\be 
E_s =  \frac{\hbar^2 \pi^2}{2 m a^2} -
\frac{e^2}{\epsilon \sqrt{2} b } + E_{ion}. \label{eq:E1d} 
\ee

At first glance it seems natural to think that when
the voltage $V$ is increased to $eV_s = E_s$ the first electron-cation pair enters
an empty SC. However, this is incorrect, as can be seen by considering the total energy of the $\nu = 1$ state, at which each
NC of the SC is occupied by a single electron that is compensated by a neighboring cation.  In this way the whole rock-salt lattice
of singly charged NCs and cations is formed. The 
electrostatic energy per pair in this lattice is actually
\be 
E_{M_1} = - M_1 \frac{e^2} { \epsilon \sqrt{2} b},
\label{eq:Em1} 
\ee
where $M_1 \simeq 1.75 $ is the Madelung constant for a rock-salt
lattice \cite{madelung}. Accordingly, the total energy per
electron-cation pair in the filled SC lattice is
\be 
E_{\nu = 1} = E_q + E_{M_1} + E_{ion} = \frac{\hbar^2 \pi^2}{2 m a^2} -
1.75 \frac{e^2}{\epsilon \sqrt{2} b} + E_{ion}. 
\label{eq:Ed1} 
\ee

Comparing Eq.\ (\ref{eq:Ed1}) to Eq.\ (\ref{eq:E1d}) we see that for the fully filled lattice
the energy per pair is
smaller by $ 0.51 e^2/\epsilon b$. Therefore, when the voltage $V$
is tuned up, no electron-cation pairs enter the empty SC
until the critical voltage 
\be 
e V_1 = E_{\nu =1} =
\frac{\hbar^2 \pi^2}{2 m a^2} - 1.75 \frac{e^2}{\epsilon \sqrt{2}b} + E_{ion},
\label{eq:Vg1dp} 
\ee 
at which $N$ such pairs enter collectively.
Thus, charging of the SC happens as a first order phase transition.
As a consequence, the capacitance has a $\delta$-function peak at $V = V_1$
(Fig.\ \ref{fig:charcap}). 

One can also use the language of dipole moments formed by nearest
neighboring electron-cation pairs in the SC-RTIL capacitor to illustrate
the mechanism of this first order transition. In the rock salt arrangement
of electrons and cations, apparently the repulsion between neighboring, aligned dipoles is overcome by the attraction between anti-aligned dipoles (Fig.\ \ref{fig:lattice}). This happens because the vector connecting the centers of nearest neighboring dipoles forms an angle of $\pi/4$ with respect to the dipole moment. In this sense, the SC-RTIL
capacitor in the form of a rock-salt lattice of electrons and
cations is dramatically different from a more conventional capacitor
where all dipoles repel each other (Fig.\ \ref{fig:standardcapacitor}). It is the dipole-dipole attraction that leads
to cooperativity in the charging of the SC-RTIL capacitor.

For the SC of small NCs discussed above, the cations can fit into the octahedral pores only. However, for a SC of larger NCs, where the pore size is thereby larger, the lattice can accommodate cations in its tetrahedral pores as well and a second delta peak of capacitance might occur. We show below that, for the case of cations residing in the tetrahedral pores, the system can exhibit cooperativity at $\nu = 1$ and $\nu = 2$ only. 

Recall that there are two tetrahedral pores per NC within an fcc lattice. For singly charged NCs, the cations fill half (one sublattice) of the tetrahedral pores instead of the octahedral ones. The corresponding lattice of charges resembles the zinc blende lattice such as ZnS \cite{crystal}, where each NC takes the place of S$^{2-}$ and cations occupy Zn$^{2+}$ sites. To bring a single electron-cation pair into the empty SC, it costs energy 
\be 
E_s = \frac{\hbar^2 \pi^2}{2 m a^2} -
\frac{e^2}{\epsilon \frac{\sqrt{6}}{2} b } + E_{ion},
\label{eq:Est} 
\ee
where the first term is the electron ground state energy $E_q$, the second term represents the Coulomb interaction between the electron and its nearest neighboring cation ($\sqrt{6}/2 b$ is the distance between them), and $E_{ion}$ is the chemical potential of cations in the bulk RTIL. For the fully filled SC where each NC is occupied by an electron, the total energy per electron-cation pair is  
\be 
E_{\nu = 1} = E_q + E_{M_2} + E_{ion} = \frac{\hbar^2 \pi^2}{2 m a^2} -
1.64 \frac{e^2}{\epsilon \frac{\sqrt{6}}{2} b} + E_{ion}. 
\label{eq:Eft1} 
\ee
where $1.64$ is the Madelung constant for a zinc blende lattice \cite{madelung}. 

Comparing Eq.\ (\ref{eq:Est}) with Eq.\ (\ref{eq:Eft1}) we find that energy per pair is smaller by $0.52 e^2/\epsilon b$, which means that the first electron enters each NC simultaneously at $eV_1 = E_{\nu = 1}$, and that at $V_1$, the capacitance $C(V_1)$ shows a delta peak (Fig.\ \ref{fig:charcap}). Comparing Eq.\ (\ref{eq:Eft1}) with Eq.\ (\ref{eq:Ed1}), we see that a cation indeed prefers a tetrahedral pore to an octahedral one. The reason is that in the former pore the cation is closer to its neighboring electron than in the latter. 

When a second electron enters each NC, it occupies the same $1s$ state but with opposite spin. To make the assembly overall neutral, cations from the bulk RTIL fill the other half of the tetrahedral pores so that there are two cations per NC. This lattice of point charges is equivalent to the fluorite CaF$_2$ lattice. Each NC with two electrons takes the place of Ca$^{2+}$ and cations occupy F$^-$ sites. 

To see whether the system exhibits cooperativity at $\nu =2$, one should compare the energy of a single additional electron-cation pair with the energy per pair of the state where $N$ additional electrons enter the zinc blende lattice collectively. When only one additional electron enters, a cation resides in the originally empty neighboring tetrahedral pore to compensate for it. The energy cost to bring this pair into the zinc blende lattice is therefore
\be 
E_s = E_q + E_r + E_{ion} + E_c - \frac{e^2}{\epsilon \frac{\sqrt{6}}{2} b}
\label{eq:Est2} 
\ee
where $E_r$ is the repulsion between two electrons within a NC, $E_c + E_r$ is the electrostatic interaction energy between the zinc blende lattice of singly charged NCs and cations in first sublattice of tetrahedral pores and the newly-inserted nearest neighboring electron-cation pair, the last term is the Coulomb interaction between the pair, and $E_q$ and $E_{ion}$ are defined in the same way as above. In contrast, when $N$ additional pairs enter the SC, they form a second zinc blende lattice and the total energy per such pair is as follows:
\be 
E_{\nu = 2} - E_{\nu = 1} = E_q + E_r + E_{ion} + E_c - 1.64 \frac{e^2}{\epsilon \frac{\sqrt{6}}{2} b}
\label{eq:Eft2}. 
\ee
By comparing Eq.\ (\ref{eq:Est2}) with Eq.\ (\ref{eq:Eft2}) we see that again the energy per additional pair is lower than that of a single additional pair. This means that cooperativity exists at $\nu = 2$.

Since $E_c$ is unknown, to obtain the voltage $V_2$ at which a second electron enters each NC simultaneoulsy, let us first calculate the total energy per pair of the SC at $\nu = 2$. This energy is the sum of the quantum mechanical energy $2 E_q$, the direct Coulomb repulsion $E_r$ between two electrons within a NC, the electrostatic energy per pair of two electrons and two cations $E_{M_3}$, and the chemical potential of the cations $2 E_{ion}$:
\begin{eqnarray}
E_{\nu = 2} & = & 2 E_q + E_r + E_{M_3} + 2 E_{ion} \\ \nonumber
& = & \frac{\hbar^2 \pi^2}{m a^2} + 1.79 \frac{e^2}{\epsilon a} - 5.04 \frac{e^2}{\epsilon \frac{\sqrt{6}}{2} b} + 2 E_{ion}.
\label{eq:Ed2} 
\end{eqnarray}
Here, $ 5.04 $ is the Madelung constant for a fluorite lattice \cite{madelung}. Therefore,
\be
e V_2 = E_{\nu = 2} - E_{\nu = 1} = \frac{\hbar^2 \pi^2}{2m a^2} + 1.79 \frac{e^2}{\epsilon a} -3.4 \frac{e^2}{\epsilon \frac{\sqrt{6}}{2} b} + E_{ion},
\label{eq:Vg2d} 
\ee
where a second delta-function of capacitance is located at  potential $V_2$. 

For higher filling factor like $\nu = 3$, since both $1s$ states are already occupied, a third electron entering NC will be in the $1p$ state. This quantum mechanical energy is greater than the Coulomb interaction of electron-cation pairs and thereby causes a large gap between $V_2$ and $V_3$. This gap also increases because one can expect larger energy necessary to bring a third cation per NC into the assembly. To take into account this effect one should change the model and go beyond one cation per pore or consider SC lattice swelling.  We have not tried to explore such more complicated models yet. 

\label{sec:semi}

\section{Metallic Nanocrystal Assembly}

Throughout Sec. III, the dielectric constant $\epsilon$ was regarded
as uniform over the whole lattice.  This is an oversimplification
of a real system, since semiconductor NCs usually
have $\epsilon$ larger than that of a RTIL. As a consequence, one can expect that cations
create image charges in surrounding NCs, which complicates the calculation of energy as a function of cation concentration. To understand whether image charges can qualitatively
affect cooperativity, in this section we study metallic NCs, which corresponds to the case where NCs have infinite dielectric constant, where the effect of images should
be even stronger.  Specifically, we consider a SC of electrically isolated spherical metallic NCs with external radius $b$ (see Fig.\ \ref{fig:lattice}).

Suppose that the insulating film covering NCs makes the overlap between
electron wavefunctions of neighboring metallic NCs so weak that the
charge on each NC is quantized and the SC conducts only by electron hopping.
Metallic NCs are assumed to form a face-centered cubic lattice
and cations reside in the octahedral pores (Fig.\ \ref{fig:device}). We show below that for this system
the cooperativity of electrons still exists for the case of metal NCs
in the vicinity of filling factor $\nu = 1$. This conclusion suggests that
a SC of semiconductor NCs (with large but finite dielectric constant) gated by RTIL should also show cooperativity
near filling factor $\nu = 1$.

Let us first calculate the electrostatic energy of the system occupied by a single pair of a cation and an electron in its adjacent NC. By classical electrostatics we know that the cation will induce images in all metal NCs of the SC \cite{LandauLifshitz}. These image charges produce new images in every other metal sphere, and so forth. Thus we have an infinite series of image charges in each sphere. The problem of image charge summation has previously been discussed for the case of a point charge in the vicinity of two conducting spheres \cite{image}; here we discuss the problem of an infinite lattice of conducting spheres.

The electrostatic energy cost to bring a single electron-cation pair into the empty lattice is determined by the total interaction energy of the cation with all the induced images.  It should be noted that for a cation in the center of an octahedral pore the charges created in the cation's six nearest neighboring NCs are dominating. We computed the electrostatic interaction energy between one cation and the image charges induced in its six nearest neighboring NCs. To solve this problem numerically, we construct an infinite series of image charges via a recursion process (see Appendix A). The computational result shows that the electrostatic energy of an electron-cation pair in the empty lattice is approximately $ E_{s} = (-1.06 \pm 0.1) e^2/b$, where $b$ is the external radius of NC.

We now turn to the calculation of the energy per pair in a fcc lattice of singly charged metal NCs with cations in all the octahedral pores. Since each cation creates images in every metal sphere of the SC, the image charge summation becomes very difficult to carry out. 
In order to solve this problem we adopt the following numerical strategy.  We first replace each cation by a metallic sphere with vanishingly small radius $a_{c}$ and charge $+e$.  We then numerically solve the Laplace equation at a given V by the
Gauss relaxation method \cite{relaxation}.  From this numeric solution one can determine the charging energy, namely the total energy of the system per electron-cation pair (see Appendix B for more details). To obtain the interaction energy per such pair we need to subtract the self-energy of a metal sphere $e^2/2b$ and of a cation $e^2/2a_{c}$ in isolation from the charging. Numerical results give $ E_{\nu = 1} = (-1.48 \pm 0.01) e^2/b $ at $a_{c}\ll b$.

Comparing $E_s$ with $E_{\nu = 1}$, we see that the electrostatic energy per pair in the filled lattice is lower than that of a single pair, which means that there is indeed cooperativity of electrons in a filled metallic SC. This result can be directly applied to a SC of heavily doped semiconductor NCs, which behave as metallic dots. On the other hand, it also implies that for a SC of undoped semiconductors the large dielectric constant of the semiconductor does not suppress cooperativity in the charging process of a SC gated by RTIL. 

In order to test the generality of these conclusions, we examined the case in which metallic NCs form a simple cubic lattice instead of an fcc lattice. For a simple cubic lattice, each unit cell contains one cation residing in the center and eight nearest neighboring NCs occupying the corners.  The methods described in Appendix A and B are employed to numerically compute the electrostatic energy of a single electron-cation pair and the energy per pair in a filled lattice. As expected, the simple cubic lattice of metallic NCs exhibits cooperativity as well: the electrostatic energy per pair of a fully-filled lattice is approximately $-1.22e^2/b$, lower than that of the system occupied by a single pair, which is about $-0.63e^2/b$. This again corroborates our prediction that cooperativity is a general property of electrons in nanocrystal SCs rather than a phenomenon applicable only to a few special cases. A more general proof of cooperativity of SC of metallic NCs is given in Appendix C.

Until now we have not discussed the difference between the charge distributions in metallic NCs and semiconductor NCs with large dielectric constant $\epsilon$. For metallic NCs, all charges are distributed on the surface, while for semiconductor NCs, the injected electron is confined within the NC with a ground state wave function given by Eq.\ (\ref{eq:wavefunction}). To see whether this difference affects cooperativity, let us first look at a singly charged NC and its adjacent cation. For the case of a metallic NC, the surface charge of the NC can be seen as a superposition of two parts: (1) charge that is induced by the cation in a grounded NC, and (2) an electron charge $-e$ that is uniformly distributed on the surface. For a semiconductor NC, the total charge is also composed of two parts, but the second part is an electron with the spherically symmetric ground state wave function given by Eq.\ (\ref{eq:wavefunction}). In the semiconductor case one should pay an additional self energy $E_{as}$ to assemble the electron distribution implied by Eq.\ (\ref{eq:wavefunction}) from the uniform surface charge which has total charge $-e$. $E_{as}$ is the same for each NC and therefore is independent of whether the lattice is empty or filled. When considering a fully filled lattice of semiconductor NCs, since every NC has such an energy difference due to charge redistribution, the critical voltage at which cooperativity shows up is in fact shifted by $E_{as}/e$ compared to the result of metallic NCs. Apparently this shift of threshold voltage does not have any influence on cooperativity. Therefore, our analogy between metallic NCs and semiconductor NCs with large dielectric constant is conclusive. 

\label{sec:metal}

\section{Conclusion}

In summary, we have shown that the capacitance of an ideal supercrystal of semiconductor NCs gated by a RTIL is the sum of delta-functions located at critical gate voltages. At each such critical voltage every NC aquires an additional electron. This was shown assuming that cations arrive in the middle of the pores of the supercrystal and arrange themselves in a crystalline lattice.
To our mind this is an interesting theoretical model of a macroscopic system of interacting nanoparticles which demonstrates the capacitance behavior expected for an ensemble of non-interacting quantum dots. 

In our theory we ignored the effects of finite temperature and disorder on the crystalline
lattice of cations and charged NCs. The thermal energy $k_BT_R$ corresponding to room temperature $T_R$ should be compared with the gain which a nearest neighbor pair of a charged NC and a cation gets when cations and electrons enter cooperatively. This gain $W \simeq 0.5e^2/\epsilon b$, as we saw above for the system with uniform dielectric constant. For a semiconductor NC with dielectric constant much larger than the dielectric constant $\epsilon_{i} \sim 2$ of the RTIL, the energy $W \simeq 0.5e^2/\epsilon_i b \sim 5k_BT_R$ for $b =2.5$ nm. This means that temperature plays a secondary role in the thermodynamics of our system and the capacitance still should have delta function peaks. 

Now let us discuss the role of a spatial disorder. One of the sources of such disorder can be size fluctuations $\delta a$ of the NC radius $a$. In experiment the relative fluctuations  $\delta a /a$ are typically around $7\%$~\cite{Heng}. These fluctuations of course affect the structure of the SC and all Coulomb energies. But for relatively small NCs the most important role of size fluctuations is the variation of the large quantum mechanical energy of electrons within the NC.  
For $a = 2$ nm and relative size fluctuation = $7\%$, for example, we get 
\be
\delta E_q  = E_q \frac{2 \delta a}{a} =  0.14 E_q \simeq 0.1 eV = 4k_{B}T_R,
\label{eq:Eqmf} 
\ee
Such fluctuations are comparable with the energy gain associated with cooperativity and may destroy cooperativity effects and smear out the delta peaks of capacitance predicted above. Instead, if smaller fluctuations of size of order of $4\%$~\cite{Murray2} are achievable one can count on the observation of delta function peaks in the capacitance. However, we are not aware of any experimental demonstration of such a capacitance. 

The most advanced study of charging of a SC of almost identical semiconductor NCs~\cite{Heng} focuses on the dependence of conductivity of the SC, $\sigma(T)$, on the gate voltage $V$. The authors discovered that the conductivity follows the variable range hopping (VRH) law $\sigma(T)\propto
\exp[-(T_0/T)^{s}]$, where the power $s$ oscillates with growing $V$ from $s=1/4$ (Mott VRH~\cite{Mott}) to  $s=1/2$ (Efros-Shklovskii VRH~\cite{ES1975}).
(Such periodic oscillations were predicted in Ref.~\cite{Jingshan} for periodic granular system of metallic dots doped by donors that are randomly positioned in the spaces between the dots.) Apparently, observation of variable range conductivity in the samples of Ref.~\cite{Heng} points to the important role of disorder. This disorder may be a random distribution in sizes, which leads to fluctuations in the quantum energy, as discussed above. One can also imagine that 
in a system with narrow pores cations never become completely equilibrated to their crystalline ground state arrangement.  These issues should all be clarified by future experiments, where capacitance data should be analyzed together with transport one.

\vspace*{2ex} \par \noindent
{\em Acknowledgments.}

We are grateful to Alexander Efros and P.\ Guyot-Sionnest for helpful discussions.  This work was supported partially by the MRSEC Program of the National Science Foundation under Award Number DMR-0819885.  T.\ C.\ thanks the FTPI for financial support.

\vspace*{2ex} \par \noindent
{\textbf {Appendix A: Image Charge Summation}}
\vspace*{2ex}

In this Appendix we show how an infinite series of image charges is 
formed in each metal sphere of the SC surrounding a point charge. Consider a cation in the octahedral pore 
surrounded by six metal spheres (metal NCs). Let us take the singly 
charged NC adjacent to the cation as an example. The cation first creates 
an image charge $- e' = - e b/\sqrt{2} b = - e/\sqrt{2}$ ($\sqrt{2}b$ is 
the nearest neighbor distance) lying on the ray from the center of a 
neighboring NC to the cation, as described in Ref.\ 
~\cite{LandauLifshitz}. To maintain the total charge $-e$ in the metal 
sphere and equipotential on the surface, a second image $e'' = -e + 
e/\sqrt{2}$ has to be placed at the center ~\cite{LandauLifshitz}. These 
two charges induce images respectively inside all the other spheres. The 
newly induced images in each neutral NC will again create new images in 
the singly charged NC, and so forth. Meanwhile, the same process is 
repeated for all the other NCs, except that the total charge in each of 
them is zero instead of $-e$.  In this way an infinite series of image 
charges is formed inside every metal sphere.

The above image generating process can be written as a simple recursive 
list of rules:

1) Create the image charges $-e'$, $-(e-e')$ in the singly charged metal 
sphere.

2) Create the image charges $-e'$, $+e'$ in the other five neutral spheres.

3) Create images within every sphere for all image charges inside the 
other spheres.

4) Add an additional charge to the center of every sphere to ensure that 
it has net charge $-e'$ (singly charged NC) or $0$ (neutral NC).

5) Repeat steps 3) - 4).

The total interaction energy of the cation with the surrounding NCs can 
be found by summing up the Coulomb interaction energy of the cation with 
all the induced images.  The total energy of the system occupied by a 
single pair is $1/2$ times this value. (Note that the factor $1/2$ for 
the interaction energy arises because of the image charge 
\cite{LandauLifshitz}.)

It should be noted that when a large number of metal spheres are included in the region surrounding the point charge, the above recursion process does not necessarily converge as a function of iteration number.  (Since the number of image charges created during each recursion step grows exponentially as a function of recursion number, and since these image charges are predominantly of the same sign during each step, the calculated energy oscillates as a function of iteration number and does not necessarily converge.)  This problem can be circumvented if one introduces an artificial damping of the magnitude of the image charges created during step 3), such that image charges created during later recursion steps have progressively smaller magnitude.  The results for electrostatic energy presented in Sec.\ \ref{sec:metal} correspond to a damping of the image charges by a factor $\eta^K$, where $K$ is the iteration number and $\eta$ is some positive number satisfying $\eta < 1$ and $1 - \eta \ll 1$.  For the six neighboring spheres of the fcc lattice, as considered in Sec.\ \ref{sec:metal}, this damping is sufficient to produce convergence when $\eta \lesssim 0.95$.  The convergent result for energy, $E_s = (-1.06 \pm 0.1)e^2/b$, is seen to be independent of the choice of $\eta$ to within the stated accuracy over the range $0.65 < \eta < 0.95$.

\vspace*{2ex} \par \noindent
{\textbf {Appendix B: Gaussian Relaxation Method}}
\vspace*{2ex}

In the Gaussian relaxation method, described briefly in Sec.\ 
\ref{sec:metal}, we replace each cation by a metallic sphere with small 
radius $a_{c}$ (usually $\sim 15\%$ of $a$) and charge $+e$. This charge 
is supported by an unknown voltage $V_e$ that is applied between the 
cations and the metal NCs. That is, all cations are held at $V_e$ while 
all metal spheres are held at zero potential. Since initially the value of $V_e$ that 
corresponds to charges $\pm e$ is unknown, an arbitrary voltage $V$ 
(which might not correspond to $\pm e$) is assigned to each cation.  We 
then numerically solve the Laplace equation at $V$ and obtain the electric 
potential distribution in the space outside small and large spheres 
using the
Gauss relaxation method. From this numeric solution one can find 
cation's charge $q$ that corresponds to $V$, and the system's capacitance 
$C = q/V$. The total energy of the system per pair is essentially the 
charging energy $e^2/2C$. Note that it is not necessary to solve $V_e$ 
since C is linear, i.e. $C = q/V = e/V_e$. This procedure can be 
summarized by the following list of steps:

1) Create a 3D lattice of grid points filling the unit cell of the periodic 
SC lattice.

2) Enforce the boundary conditions that all lattice points inside a 
metal sphere have potential $\Phi = 0$ and all points inside a cation 
have $\Phi = V$.

3) Solve for the potential at all other grid points using the Gauss 
relaxation algorithm. Enforce periodic boundaries at the unit cell edges.

4) Measure the charge $q$ of the ``capacitor" comprised of the small 
(cation) spheres and the larger (NC) spheres by constructing a Gaussian 
surface around one of the cations and calculating $ \oint \vec{E} \cdot 
d\vec{A} = 4\pi q$. The capacitance is $C = q/V$. From C we calculate 
the total energy per pair $e^2/2C$.

\vspace*{2ex} \par \noindent
{\textbf{Appendix C: Proof of Cooperativity for A Metallic SC.}}
\vspace*{2ex}

In the Sec.\ \ref{sec:metal}, we discussed a densely packed metallic SC that 
shows cooperativity at $\nu =1$. In that case, the Gauss relaxation 
method is employed to calculate the energy per electron-cation pair of a 
fully-occupied lattice, which is proven to be lower than that of a 
single pair in the system under the approximation that only the six 
nearest neighboring NCs interact with the cation. One may wonder whether 
cooperativity is a general property regardless of this approximation. To 
answer this question, in this section we would like to propose a proof 
that shows the existence of cooperativity for face-centered and cubic 
SC lattices of metallic NCs with large radius $b$ having small metallic 
spheres in their pores (Fig.\ \ref{fig:device}). We suppose that the 
whole system is neutral and all charges are discrete,
and in the ground state of the fully filled lattice all small spheres have
charge $+e$ and all large spheres have charge $-e$. Our goal is to prove 
in a general way that the energy per pair of a fully-occupied lattice, 
$u_\textrm{full}$ is lower than the energy of the system when it is 
occupied by a single $+$/$-$ pair, $u_s$. The case of point charges next 
to metal spheres can be recovered by making radius of the small spheres 
vanishingly small.

In general, the total electrostatic energy $U$ of a system of charged, 
conducting spheres is
\be
U = \frac12 \sum_{i} \sum_{j} C^{-1}_{i j} q_i q_j,
\label{eq:totalU}
\ee
where the indices $i$ and $j$ label all spheres in the system, $q_i$ is 
the charge of the sphere $i$, and $C^{-1}_{i j}$ is the inverse of the 
matrix of electrostatic induction $C_{i j}$ (this matrix has the property $C_{i j}=C_{j i}$) 
\cite{LandauLifshitz}.  This expression for energy includes all binding 
energies as well as the self-energies of all charges.

Consider first the case where only a single $+$/$-$ pair is charged: say 
that a small sphere (cation) $i = 1$ has charge $q_1 = e$ and a 
neighboring large sphere (metal NC) $i = 2$ has charge $q_2 = -e$, while 
all other spheres have $q_i = 0$.  Then the double sum in Eq.\ 
(\ref{eq:totalU}) is reduced to only four terms, and the total energy 
$u_s$ becomes
\be
u_s = \frac{e^2}{2} \left[ C^{-1}_{11} + C^{-1}_{22} - 2C^{-1}_{12} \right].
\label{eq:u1}
\ee

Now consider the case where the lattice is entirely filled: all the 
small spheres (cations) in the lattice have $q = e$ and all the large 
spheres (NCs) have $q = -e$.  Suppose, further, that the lattice is in 
its lowest energy configuration.  This implies that the system must be 
stable with respect to exchanging the charges of any two spheres.  That 
is, if the sign of the charge at site $i$ and site $j$ are 
simultaneously inverted, then the system energy must increase or be 
unchanged.

Suppose, without loss of generality, that in the ground state of the 
filled lattice the site $i = 1$ has charge $+e$ and the site $i = 2$ has 
charge $-e$.  From Eq.\ (\ref{eq:totalU}), one can therefore write the 
system's initial energy as

\begin{eqnarray}
U_{in} & = & \frac12 \left[ e^2 C^{-1}_{11} + e^2 C^{-1}_{22} -2 e^2 
C^{-1}_{12} \right] \\  \nonumber
& & + \frac12 \left[ 2 \sum_{j \neq 1,2} (e q_j C^{-1}_{1j} - e q_j 
C^{-1}_{2j} )\right] \\ \nonumber
& & + \frac12 \left[ \sum_{i \neq 1,2} \sum_{j \neq 1,2} C^{-1}_{ij} q_i 
q_j  \right].
\label{eq:Ui}
\end{eqnarray}

If the charges in site $i = 1$ and $i = 2$ are exchanged, then the 
energy becomes
\begin{eqnarray}
U_{fin} & = & \frac12 \left[ e^2 C^{-1}_{11} + e^2 C^{-1}_{22} -2 e^2 
C^{-1}_{12} \right] \\ \nonumber
& & - \frac12 \left[ 2 \sum_{j \neq 1,2} (e q_j C^{-1}_{1j} - e q_j 
C^{-1}_{2j} ) \right] \\ \nonumber
& & + \frac12 \left[ \sum_{i \neq 1,2} \sum_{j \neq 1,2} C^{-1}_{ij} q_i 
q_j \right].
\label{eq:Uf}
\end{eqnarray}

The stability of the ground state requires that $U_{fin} \geq U_{in}$.  
By comparing $U_{in}$ and $U_{fin}$, one can see that this condition implies

\be
\sum_{j \neq 1,2} (e q_j C^{-1}_{1j} - e q_j C^{-1}_{2j} ) \leq 0.
\label{eq:Uleq}
\ee

When the lattice is fully filled, the energy per pair $u_\textrm{full}$ 
can be found by taking the sum in Eq.\ (\ref{eq:totalU}) over $i = 1,2$ 
only, since $i = 1,2$ labels a single nearest-neighbor pair that is 
equivalent to all other pairs:
\begin{eqnarray}
u_\textrm{full} & = & \frac12 \sum_{i=1,2} \sum_{j} C^{-1}_{i j} q_i q_j 
\nonumber \\
& = & \frac12 \left[ e^2 C^{-1}_{11} + e^2 C^{-1}_{22} - 2 e^2 
C^{-1}_{12}\right] \label{eq:ufullu1} \\
& & + \frac12 \left[\sum_{j \neq 1,2} (eq_j C^{-1}_{1j} - eq_j 
C^{-1}_{2j} ) \right]. \nonumber
\end{eqnarray}
By substituting Eqs.\ (\ref{eq:u1}) and (\ref{eq:Uleq}) into Eq.\ 
(\ref{eq:ufullu1}), we can conclude that
\be
u_\textrm{full} \leq u_s,
\ee
so that the lattice is proven to exhibit cooperativity. For the case of 
a metallic SC with insulating cationic charges, one can simply take the 
limit where the radius of small (positive) spheres is vanishingly small, 
and one recovers the case of point charges next to metallic spheres.


\end{document}